\documentclass{osa-article}
\journal{oe}
\usepackage{graphicx} 
\usepackage{xcolor} 

\begin{document}

\title{Cherenkov Radiation Induced by Megavolt X-Ray Beams in the Second Near-Infrared Window}

\author{Xiangxi Meng,\authormark{1,2} Yi Du,\authormark{3} Ziyuan Li,\authormark{1} Sihao Zhu, \authormark{1} Hao Wu,\authormark{3,7} Changhui Li, \authormark{1,8} Weiqiang Chen, \authormark{4} Shuming Nie, \authormark{5} Qiushi Ren \authormark{1} and Yanye Lu \authormark{6,9}}

\address{\authormark{1}Department of Biomedical Engineering, College of Engineering, Peking University, Beijing 100871, China\\
\authormark{2}The Wallace H. Coulter Department of Biomedical Engineering, Georgia Institute of Technology \& Emory University, Atlanta, GA 30332, USA\\
\authormark{3}Key laboratory of Carcinogenesis and Translational Research (Ministry of Education/Beijing), Department of Radiation Oncology, Peking University Cancer Hospital \& Institute, Beijing 100142, China\\
\authormark{4}Institute of Modern Physics, Chinese Academy of Sciences, Key Laboratory of Heavy Ion Radiation Biology and Medicine of Chinese Academy of Sciences, Key Laboratory of Basic Research on Heavy Ion Radiation Application in Medicine, Gansu Province, Lanzhou 730000, China\\
\authormark{5}Department of Biomedical Engineering, University of Illinois at Urbana-Champaign, Urbana, IL 61801, USA\\
\authormark{6}Pattern Recognition Lab, Friedrich-Alexander-University Erlangen-Nuremberg, 91058 Erlangen, Germany\\
\authormark{7}13552661030@139.com\\
\authormark{8}chli@pku.edu.cn\\
\authormark{9}yanye.lu@pku.edu.cn}

    \begin{abstract*}
        Although the Cherenkov light contains mostly short-wavelength components, it is beneficial in the aspect of imaging to visualize it in the second near-infrared (NIR-II) window. In this study, Cherenkov imaging was performed within the NIR-II range on megavolt X-ray beams delivered by a medical linear accelerator. A shielding system was used to reduce the noises of the NIR-II image, enabling high quality signal acquisition. It was demonstrated that the NIR-II Cherenkov imaging is potentially a tool for radiotherapy dosimetry, and correlates well with different parameters. The NIR-II Cherenkov imaging is less susceptible to scattering, while more susceptible to absorption, compared with the visible-near-infrared imaging. Finally, a mouse was used to demonstrate this technology on animals. These results indicate the potentials to apply NIR-II Cherenkov imaging in the practice of radiotherapy.
    \end{abstract*}

    \section{Introduction}

    Precise delivery of prescribed dose to target volumes is one of the never-ending challenges in cancer radiotherapy. Due to its inherent property, ionizing radiation could only be detected indirectly through its effects. Cherenkov radiation is such an effect of the therapeutic megavolt X-ray radiation generated by the medical linear accelerator (LINAC), and it boasts many benefits in clinical dose validation during radiotherapy\cite{Axelsson_Davis_Gladstone_2011}. Clinical trials have been conducted to fully exploit these benefits in superficial dose validation of X-ray photon therapies and electron beam therapies\cite{Xie_Petroccia_Maity_2018, Jarvis_Zhang_Gladstone_2014}.

\begin{figure}[htbp]
        \centering
        \includegraphics{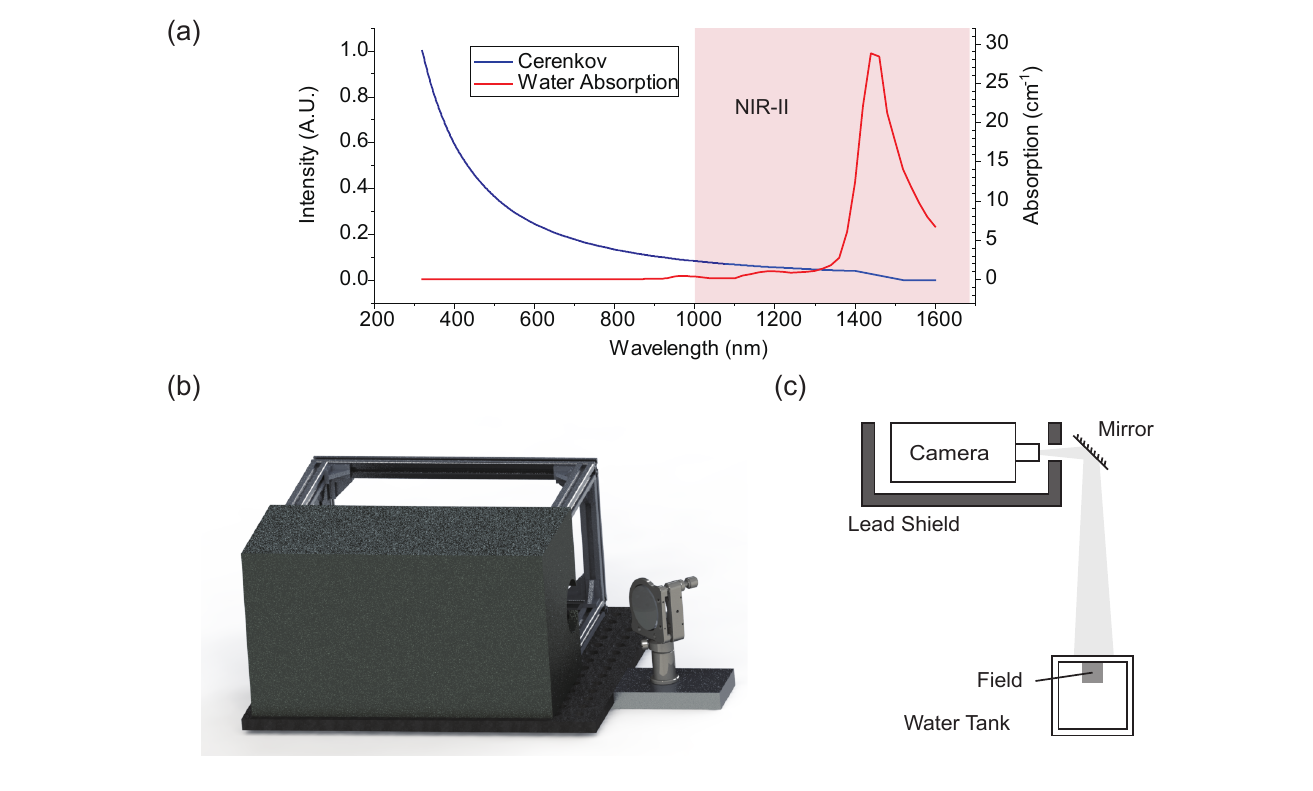}
        \caption{The Cherenkov spectrum and water absorption spectrum\cite{hale1973optical} are shown in (a). The setup of the system is shown as in (b), the rendered three-dimensional model of the shielding system and (c) the diagram showing the relative position of the equipment.\label{fig:design}}
    \end{figure}

    Cherenkov radiation is electromagnetic radiation emitted when a charged particle travels in a dielectric medium at a speed higher than the corresponding light speed in such a medium. The Cherenkov radiation is continuous in the frequency domain, and can be observed in the range from ultraviolet to infrared. As derived from the Frank-Tamm formula\cite{Frank_Tamm_1937}, the light intensity ($I_C$) generated by an elementary-charged particle is roughly proportional to the inverse square of the wavelength ($\lambda$):
    \begin{equation}
        I_C = \frac{\pi e^2 \mu}{h c \lambda^2}\left(1-\frac{c^2}{v^2n(\lambda)^2}\right)\label{eq1}
    \end{equation}
    where $e$ is the elementary charge, $\mu$ is the permeability of the medium, $h$ is the Plank constant, $c$ is the speed of light, $v$ is the speed of the electron, and $n(\lambda)$ is the wavelength-dependent refractive index. Thus, the intensity of Cherenkov radiation is higher on the blue end of the spectrum\cite{Glaser_Zhang_Davis_Gladstone_Pogue_2012}, as shown in Fig. \ref{fig:design}(a). This feature, however, limits the application of the Cherenkov radiation in both clinical radiotherapy and biomedical research, as photons with shorter wavelength generally do not propagate very deeply inside the tissue due to the elevated scattering effect. \textit{In vivo} optical imaging favors near-infrared (NIR) over visible (Vis) light, and nowadays, the second near-infrared (NIR-II) window ranging from 1000 nm to 1700 nm is attracting much attention\cite{Smith_Mancini_Nie_2009, Hong_Antaris_Dai_2017}. For instance, great success has been attained to image the NIR-II range Cherenkov photons recently, in electron beam therapy\cite{Cao_Jiang_Jia_2018}.

    The major benefit of the NIR-II region is the reduction on scattering. According to the Mie theory\cite{Staveren_Moes_van_1991}, the scattering coefficient of light is wavelength-dependent, and the longer the wavelength, the smaller the scattering coefficient is. This has been verified by numerous reports on biological tissues\cite{Bashkatov_Genina_Kochubey_Tuchin_2005,Bashkatov_Genina_Kochubey_Tuchin_2006, Du_Hu_Cariveau_Ma_Kalmus_Lu_2001} and tissue-mimicking phantoms\cite{Aernouts_Beers_Watte_Lammertyn_Saeys_2014}. Thus, the NIR-II region is ideal for conducting optical imaging in tissue. However, strong absorption of water\cite{hale1973optical} also plays a subtle role in the imaging ability of this wavelength range, as shown in Fig. \ref{fig:design}(a). With such a benefit, it is expected that the NIR-II Cherenkov imaging could be used to image the turbid medium of tissue with a better imaging performance, and expand the depth limit of optical dosimetry. 

    Although the electron beams are crucial in the treatment of several specific diseases\cite{griep1995electron}, it is widely acknowledged that megavolt X-ray photons enable the most prevalent treatment modalities in modern radiotherapies\cite{desrosiers2000150}, with its greater penetration ability. The major obstacle in implementing NIR-II imaging using megavolt X-ray beams is the intrinsic low sensitivity of the InGaAs camera for NIR-II detection versus the much higher readout noise induced by the X-ray photons\cite{Cao_Jiang_Jia_2018}. Fortunately, the scattered\cite{Nogueira_Biggs_2002} and leaked\cite{Li_Mutic_Low_2006} radiation of the megavolt X-ray beams are relatively easy to be protected from at the perpendicular position relative to the beam.

    \section{Experimental Setup}
    
    The general setup of the Cherenkov imaging using a water tank was similar to previously reported experiments\cite{Glaser_Davis_McClatchy_2013, Yamamoto_Okudaira_Kawabata_2018, Pogue_Zhang_Glaser_2017}, as shown in Fig. \ref{fig:design}(c). Megavolt X-ray beams were delivered to the water tanks, which were filled with water or a dilute solution of 20\% Intralipid\textsuperscript{\textregistered} (IL). The distance between the water tank and the camera lens was about 2 m.

    \paragraph{The shielding system and the NIR-II camera} A shielding system was designed to protect the camera from the radiation, as shown in Fig. \ref{fig:design}(b). The shielding was made of 10 mm lead plates assembled to cover the side of the NIR-II camera, and the camera was enclosed inside a structural framework. Safety space was reserved for efficient heat dissipation. A protected silver reflective mirror (GCC-102205 by Daheng New Epoch Technology, Inc.) was installed so that the camera lens could be further protected from the radiation. 
    
    The InGaAs-based NIR-II camera (NIRvana 640 by Princeton Instruments, Inc.)  was equipped with a 50 mm SWIR lens (SWIR-50 by Navitar, Inc.), and a 1064 nm long-pass filter (Semrock 1064 nm EdgeBasic\texttrademark by IDEX Health \& Science, LLC.), resulting in the detection range approximately between 1064 nm and 1700 nm.

    \paragraph{The Vis-NIR camera} A camera based on silicon-based charge-coupled device (Andor iKon-M 934 by Oxford Instruments plc) was used to image the Vis to NIR range (approximately 400 nm to 900 nm). This camera was not subjected to shielding, and it was placed adjacent to the NIR-II camera.

    \paragraph{The LINAC} The radiation research platform was built on a newly-installed Edge LINAC (Varian Medical Inc.). The LINAC is able to deliver 10 MV photons at the flattening filter free mode (10xFFF) at a maximum dose rate of 2400 MU/min. According to the machine information sheet, the flattering filter to deliver 10 MV photons is made of copper and tangent. In all cases, square fields of either 5 cm $\times$ 5 cm or 2 cm $\times$ 2 cm was used. The source surface distances were kept at 100 cm.

    \paragraph{The water tank} Water tanks in different sizes were manufactured with Super White Glass, and the transmittance of the materials were tested to be no less than 95\% in the NIR-II range.

    On this experimental platform, the following sets of experiments have been conducted:
    \begin{itemize}
        \item validation of the effectiveness of the shielding system,
        \item acquisition of representative Cherenkov images with the water tank,
        \item analysis of the signal intensity and quality in varied settings, and
        \item obtaining the Cherenkov image of an animal.
    \end{itemize}
    The detailed information about these experiments is provided in the following section.

    \section{Results and Discussion}
    
    \subsection{The effectiveness of the shielding system}

\begin{figure}[htbp]
        \centering
        \includegraphics{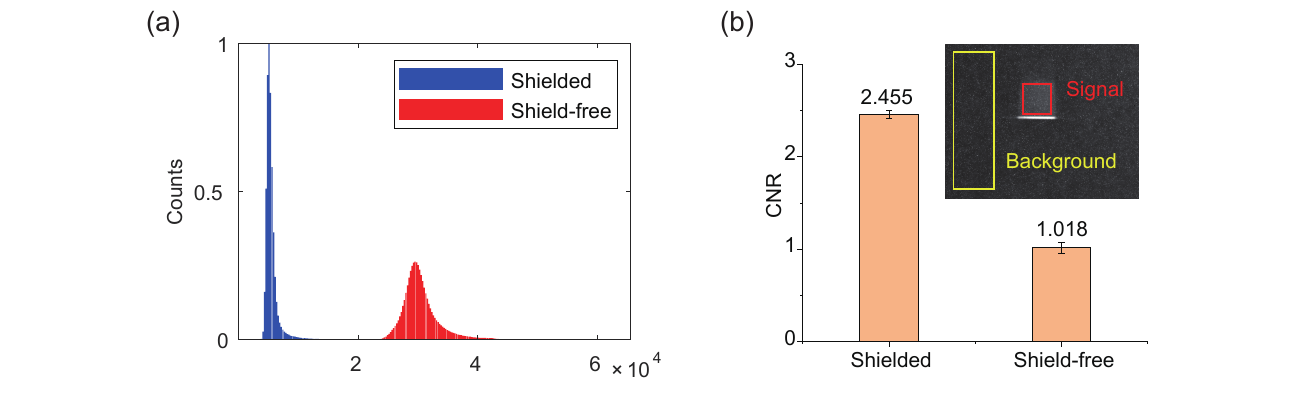}
        \caption{The performance of the Shielding system. The histograms of the dark field image acquired with and without the lead shield (a). The $\mathit{CNR}$s of a representative Cherenkov emission images with and without the lead shield (b). \label{fig:shield}}
    \end{figure}

    The effectiveness of the lead shield was first validated by comparing images acquired with or without the lead shield. Five dark field images were taken in each scenario with the LINAC delivering the same dosimetric parameters (10xFFF, 5 cm $\times$ 5 cm field, dose rate = 2400 MU/min) to a water tank. The histograms of these images were combined and compared as shown in Fig. \ref{fig:shield}(a). Then, representative Cherenkov emission images, with and without the lead shield, were taken using the same parameters. To quantify the imaging quality, the contrast-to-noise ratio ($\mathit{CNR}$) was defined as:
    \begin{eqnarray}
        C &=& S_{\mathrm{Signal}} - S_{\mathrm{Background}}\\
        \mathit{CNR} &=& \frac{C}{\sigma_{\mathrm{Background}}} 
    \end{eqnarray}
    where $C$ is the contrast of the image, $S_{\mathrm{Signal}}$ and $S_{\mathrm{Background}}$ are the average intensities of the regions of interest marked as Signal and Background, respectively, as shown in the insert of Fig. \ref{fig:shield}(b); $\sigma_{\mathrm{Background}}$ denotes the standard deviation of the intensity of the region of interest marked as Background. The $\mathit{CNR}$s of the images acquired with and without the lead shield are presented in Fig. \ref{fig:shield}(b). The effectiveness of the shielding system in noise reduction is obvious. In the dark field images, the intensity is contributed mainly by the radiation noise, and partly by system noises. When a shield was applied, the overall noise level was significantly reduced. This is also validated by the fact that the $\mathit{CNR}$ of the representative Cherenkov image increased by 2.4 fold after the shield was applied. The shielding system is crucial in reducing the noise and enabling the NIR-II Cherenkov imaging with megavolt X-ray beams.

    \subsection{The representative NIR-II Cherenkov image and analysis}

\begin{figure}[htbp]
        \centering
        \includegraphics{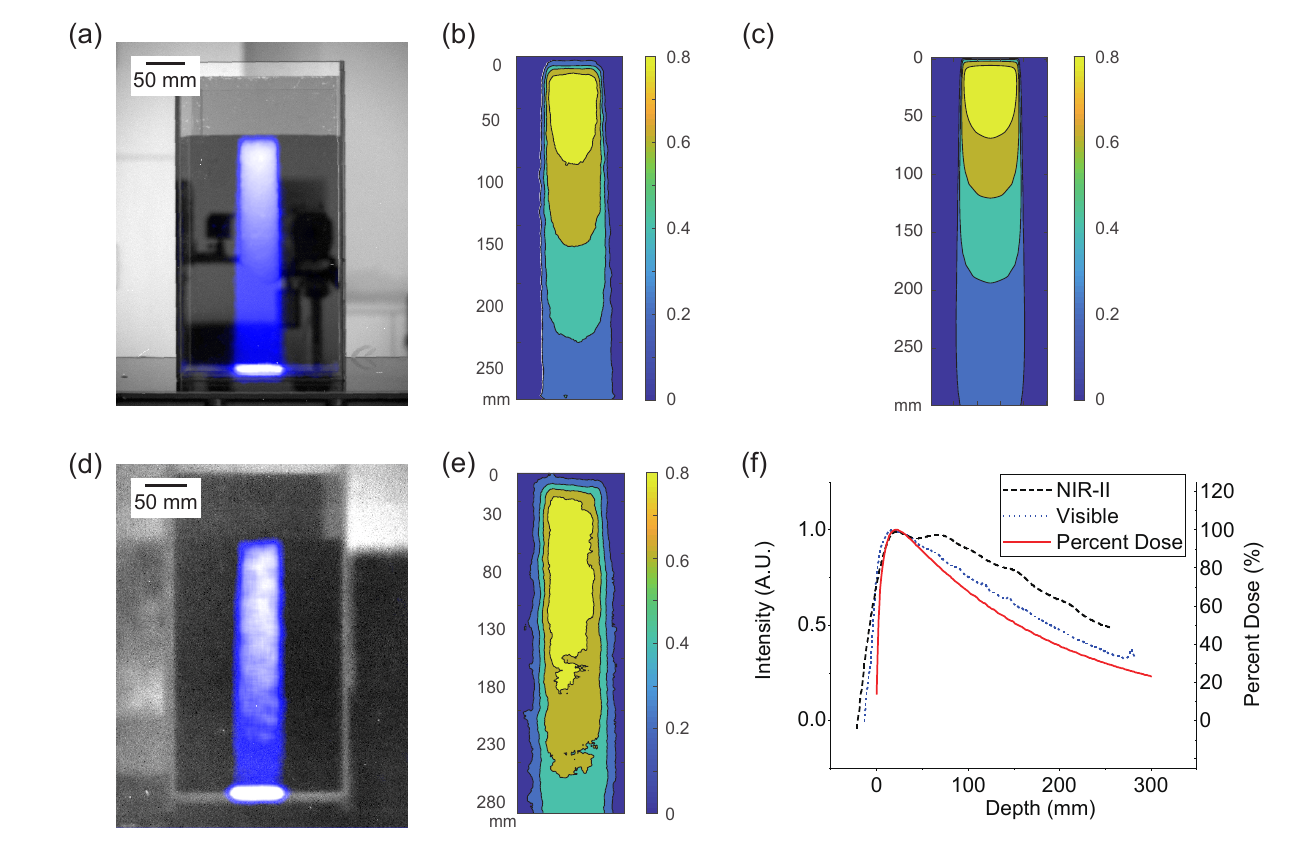}
        \caption{The overlaid image of the Cherenkov radiation in the water tank, in (a) Vis-NIR range and (d) NIR-II range. The corresponding two-dimensional Cherenkov intensity distribution in (b) Vis-NIR range and (e) NIR-II range. (c) The isodose distribution under the same dosimetric parameters. (f) The Cherenkov intensity profile along the center line and the percentage dose distribution under the same dosimetric parameters. 
        \label{fig:PDD}}
    \end{figure}

    The distribution of Cherenkov intensity, to some extent, reflects the dose distribution\cite{Glaser_Zhang_Gladstone_Pogue_2014}. Dosimetry systems have been devised to validate the superficial dose distribution based upon conventional Cherenkov imaging\cite{Brost_Watanabe_2018}. Fig. \ref{fig:PDD}(a) shows a representative Cherenkov image of a \mbox{300 mm}-deep water tank taken by the Vis-NIR camera. Contours were drawn, as shown in Fig. \ref{fig:PDD}(b), to outline the iso-intensity distribution of the Vis-NIR Cherenkov light. Using the NIR-II camera, a similar Cherenkov image in a different wavelength range can also be obtained as shown in Fig. \ref{fig:PDD}(d). The corresponding iso-intensity distribution of the NIR-II light is shown in Fig. \ref{fig:PDD}(e). We observed that the intensity distribution along the axis of the beam for the NIR-II image is more flattened than that of the Vis-NIR image. Figure \ref{fig:PDD}(c) shows the isodose distribution of the same parameters calculated by the clinical treatment planning system Eclipse (v13.6, Varian Medical Inc.), which is based on ionization chamber measurements. Although the trend is similar, it is diverted from either of the previous imaged Cherenkov intensity distribution. The intensity profile along the center line of the two dimensional iso-intensity distributions, as well as the percentage depth dose curve, are overlaid in Fig. \ref{fig:PDD}(f), showing the trend more clearly. 
    
    Hence, the NIR-II imaging is also capable of imaging the intensity distribution in the water tank. However, further investigations are required to correlate the Cherenkov intensity with the dose distribution. The unconformity between the NIR-II Cherenkov distribution and the dose distribution might be attributed to multiple causes including the anisotropy of the dependent angular distributions in different wavelength ranges of the Cherenkov spectrum \cite{Glaser_Davis_McClatchy_2013}, and the geometry aberrations of the optical systems.

    \subsection{NIR-II Cherenkov in dosimetry}

\begin{figure}[htbp]
        \centering
        \includegraphics{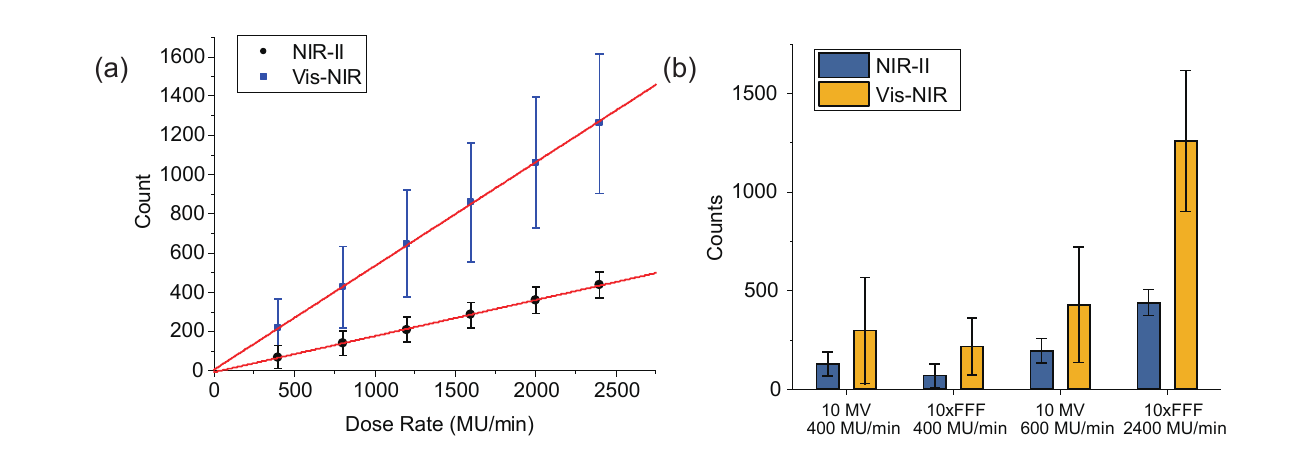}
        \caption{The relationship between the dose rate and the average Cherenkov intensity (a). The Cherenkov intensities under different parameters (b).\label{fig:Dosimetry}}
    \end{figure}

    To validate the application of the NIR-II Cherenkov imaging in dosimetry, we investigated the Cherenkov intensity in different situations, and compared them with the respective Vis-NIR Cherenkov results. Firstly, the dose rate varied from 2400 MU/min to 400 MU/min, with 10xFFF beam. The Cherenkov intensities of the NIR-II and Vis-NIR images, characterized by the average counts in the region of interest (Signal), are plotted in Fig. \ref{fig:Dosimetry}(a). The relationship between the dose rate and the intensity are both linear in the cases of Vis-NIR and NIR-II, as demonstrated by a linear regression, in which the coefficients of determination ($R^2$) being 0.9997 and 0.9991, respectively. The effect of the flattening filter was also studied. According to a recent publication\cite{Shrock_Yoon_Gunasingha_2018}, the efficiency of Cherenkov emission per dose increases when a filter is applied. Such a conclusion was validated by Vis-NIR as well as NIR-II imaging results as shown in Fig. \ref{fig:Dosimetry}(b). At the same dose rate of 400 MU/min, the intensity of Cherenkov emission of the 10 MV beam is 85\% more than that of the 10xFFF beam with the NIR-II imaging, and 37\% for Vis-NIR imaging. However, in a practical aspect, applying the flattening filter attenuates the total flux of the photon beam, thus reduces the output dose rate, which exserts a more significant effect on the Cherenkov intensity. As shown in Fig. \ref{fig:Dosimetry}(b), the maximum dose rate for 10 MV, 600 MU/min, generates a much lower Cherenkov intensity than that of the 10xFFF beam, which is 2400 MU/min.

\begin{figure}[htbp]
        \centering
        \includegraphics{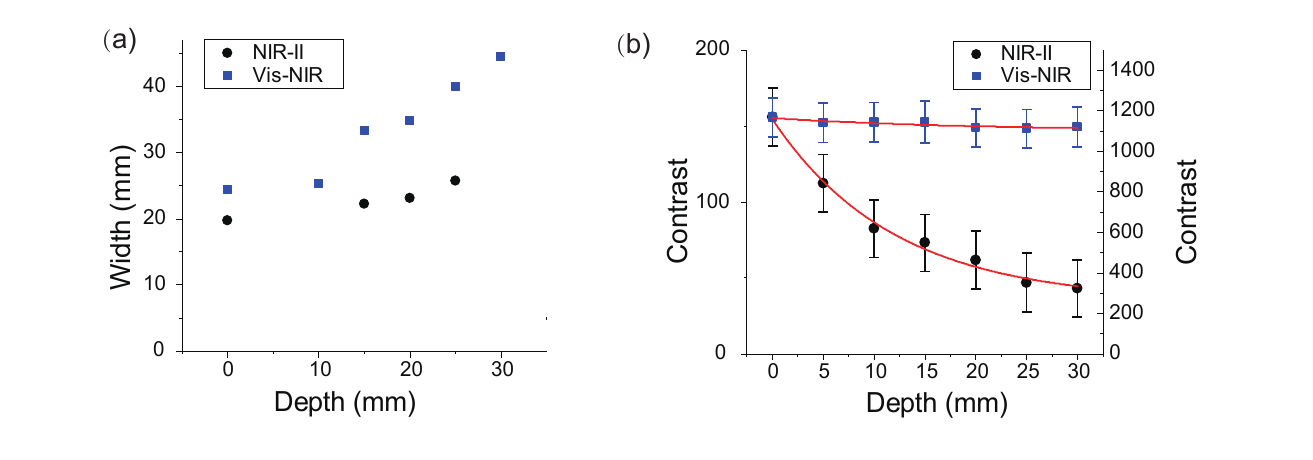}
        \caption{The average width of the Cherenkov image in different depth of IL (a). The average contrast of the Cherenkov image in deferent depth of water (b), and the error bar is 1/5 of the standard deviation, for the convenience of representation.\label{fig:IL_Ink}}
    \end{figure}

    One of the advantages of the NIR-II Cherenkov imaging in dosimetry is that, in the longer wavelength range, the scattering effect is minimized; whereas, as a trade-off, the absorption effect intensifies. Here, it was demonstrated by the depth-dependent imaging properties in diluted IL and water. In a conventional Cherenkov imaging setting, the edge of the field overlaps with the edge of the water tank. When the field moves away from the edge of the tank into the water volume, the depth is defined as the average distance between the outer edge of the field and the edge of the tank.  Fig. \ref{fig:IL_Ink}(a) shows the relationship between the depth and the average width, or full width half maximum, of the Cherenkov images in 0.5\% IL. As shown here, the width in the NIR-II range is lower than the values in the Vis-NIR range, and the trend is more obvious when the depth increases.

    Fig. \ref{fig:IL_Ink}(b) shows the relationship between the depth and the Cherenkov image contrast ($C$) in pure water. The contrast only slightly decreased in the Vis-NIR range, however, it dropped significantly in the NIR-II range. This is due to the much higher absorption coefficient of water in this range. In either case, as the scattering is negligible, the absorption follows the Beer-Lambert law
    \begin{equation}
        I(d) = I_C \exp(-\alpha d)
    \end{equation}
    where $I(d)$ is the intensity after the absorption, $d$ is the depth of the absorptive medium (water in this case), and $\alpha$ is the absorption coefficient. The data are fitted accordingly as shown in Fig. \ref{fig:IL_Ink}.

    \subsection{Animal experiment}
\begin{figure}[htbp]
        \centering
        \includegraphics{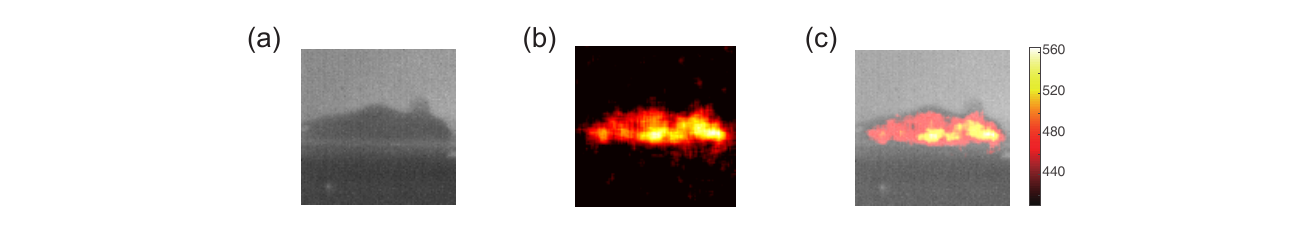}
        \caption{The white light image of a mouse (a), the Cherenkov image of the same mouse (b), and the merged image (c) .\label{fig:Mouse}}
    \end{figure}

    A balb/c nude mouse bearing a human HCT116 tumour was used in the animal study, as reviewed by the Institutional Animal Care and Use Committee of Peking University. The animal was first anesthetized and then subjected to a field that covered the whole body of it. 10xFFF beam was delivered and the animal was imaged from the side. The NIR-II Cherenkov image of the mouse with megavolt X-ray is shown in Fig. \ref{fig:Mouse}. This result proves that Cherenkov imaging with megavolt X-ray in the NIR-II range can be observed with an animal.

    \section{Conclusion}

    In this study, the Cherenkov imaging of megavolt X-ray beams was observed in the NIR-II region. This imaging method exhibits the potential of being used as a dosimetric tool for quality assurance in radiotherapy. The delivery of therapeutic dose could be monitored both qualitatively and quantitatively, enabling optical dosimetry\cite{Glaser_Zhang_Gladstone_Pogue_2014} and dynamic dose verification\cite{Black_Velten_Wang_Na_Wuu_2019}. The results obtained suggest that NIR-II imaging brings an opportunity to further improve the performance of these methods.
    
    Comparing to the conventional Vis-NIR imaging, NIR-II imaging is more capable of visualizing structured patterns through a layer of scattering medium, such as human tissue. Admittedly, NIR-II imaging suffers from a moderate absorption by water, which is a major component of the tissue. However, by applying advanced imaging technologies, the detection limit of the signal could be improved to overcome the absorption effect. In fact, water absorption may even help to increase the contrast of the NIR-II image\cite{Carr_Aellen_Franke_2018}.

    Due to the working mechanism of the LINAC, the beam delivered and the Cherenkov radiation generated are in very short pulses. This deteriorates the signal quality, thus only very low signal could be detected by the camera at a reasonable integration time. Time-gated intensified charge-coupled devices were developed to be synchronized with the beam period, and could remarkably improve the signal quality in the Vis-NIR range\cite{Demers_Davis_Zhang_2013, Andreozzi_Zhang_Glaser_2015}. However, up to now, no NIR-II camera with equivalent time-gated function is available.

    There are still many unmet scientific requirements before such technology could be fully translated into the clinical settings. Efforts should be focused on further improving the imaging quality, developing time-gated NIR-II systems, investigating on the Cherenkov-excited luminescence imaging, and data-driven image processing algorithms.

	\section*{Funding}
    The National Key Instrumentation Development Project of China (2013YQ030651); National Natural Science Foundation of China (NSFC) (81421004, 61571262, 11575095, 11505012); Natural Science Foundation of Beijing Municipality (No. 1184014, 1174016); China Postdoctoral Science Foundation (No. 2017M620542); and Capital’s Funds for Health Improvement and Research (No. 2018-4-1027)


\bibliography{reference}

\end{document}